\documentclass[12pt]{article}

\begin{document}
\begin{flushright}
hep-th/0010155
\\
AEI-2000-068
\\
MPI-MIS-65/2000
\end{flushright}

\vspace{.5cm}

\begin{center}
{\Large
An overview of new supersymmetric gauge theories\\ 
with 2-form gauge potentials
}
\end{center}

\begin{center}
Friedemann Brandt
\end{center}

\begin{center}
{\sl Max-Planck-Institut f\"ur Mathematik in den 
Natur\-wissen\-schaf\-ten,\\
Inselstra\ss e 22-26, D-04103 Leipzig, Germany\\
After October 1, 2000:\\
Max-Planck-Institut
f\"ur Gravitationsphysik, Albert-Einstein-In\-sti\-tut,\\ 
Am M\"uhlenberg 1, D-14476 Golm, Germany
}
\end{center}

\begin{abstract}
An overview of new 4d supersymmetric gauge 
theories with 2-form gauge potentials constructed by various 
authors during the past five years is given. The key r\^ole
of three particular types of interaction vertices is emphasized.
These vertices are used to develop a connecting perspective on 
the new models and to distinguish between them. 
One example is presented in detail to illustrate
characteristic features of the models. A new result
on couplings of 2-form gauge potentials to Chern-Simons 
forms is presented.
\end{abstract}

\section{Introduction}

During the past five years, several new
4d supersymmetric gauge theories have been constructed
by various authors 
\cite{bra-1}--\cite{bra-13}.
Common to all these
models is the presence of 2-form gauge potentials
and a complicated (nonpolynomial) structure of
interactions and symmetry transformations (gauge symmetries, supersymmetry).
The initial motivation to construct such models came from
string theory and focussed the attention first
on the vector-tensor (VT) multiplet
\cite{bra-VT1,bra-VT2} of N=2 supersymmetry. 
Namely, in N=2 supersymmetric 4d heterotic string vacua,
the dilaton is believed to reside in a VT multiplet
(see, e.g., section 3 of the review \cite{bra-LF}). 
In order to couple this multiplet
to N=2 supergravity, its so-called central charge must be gauged
and this leads inevitably to the structures
characteristic of the new models (cf.\ remarks at the end
of section \ref{bra-overview}).
Only two of the works \cite{bra-1}--\cite{bra-13} are not
devoted to the VT multiplet: in \cite{bra-11}
a rather general class of new supersymmetric gauge theories
with 2-form gauge fields is constructed, and \cite{bra-13}
deals with the double tensor (TT) multiplet
of N=2 supersymmetry and its couplings to vector and hyper multiplets.
The TT multiplet is believed to be the dilaton-multiplet
of N=2 supersymmetric type IIB superstring vacua \cite{bra-LF} and thus 
it should play there a r\^ole analogous to the VT multiplet 
in heterotic vacua.

The purpose of this contribution is to give an overview
of the new models and to emphasize the key r\^ole of
three types of cubic interaction vertices in these models.
To this end, first a brief
excursion to consistent interactions of $p$-form gauge potentials
in general is made in section \ref{bra-sec2}. This will also show
how the new models fit in the recent
classification \cite{bra-HK1,bra-HK2,bra-HK3} of interactions
between $p$-form gauge potentials. The
three particular types of interaction vertices are identified
and discussed in some detail in section \ref{bra-overview},
including a new result on couplings of 2-form gauge potentials
to Chern-Simons forms. Then
these vertices and the supersymmetry multiplet structure
are used to characterize the various
models and to distinguish between them.
In section \ref{bra-example}, an explicit example
is treated in detail to illustrate characteristic features 
of the new models.
The example is an N=2 supersymmetric model 
found in \cite{bra-13},
coupling the TT multiplet mentioned above
to two N=2 vector multiplets. Section \ref{bra-comments} contains 
a selection of open problems and possible
future developments.

\section{Interactions of $p$-form gauge potentials}\label{bra-sec2}

Gauge invariance restricts the possible interactions
of $p$-form gauge fields quite severely. In the simplest case,
the gauge transformation of 
a $p$-form gauge potential $A=(1/p!)
dx^{\mu_1}\wedge\dots\wedge dx^{\mu_p}A_{\mu_1\dots\mu_p}$ is a natural
generalization of the gauge transformation of the 
electromagnetic gauge field:
\begin{equation}
\delta_\mathrm{gauge}^{(0)}A=d\omega\Leftrightarrow\
\delta_\mathrm{gauge}^{(0)}A_{\mu_1\dots\mu_p}=
p\partial _{[\mu_1}\omega_{\mu_2\dots\mu_p]}\ ,
\end{equation}
where $\omega_{\mu_1\dots\mu_{p-1}}$ are arbitrary gauge
parameter fields. Analogously to
the electromagnetic case, corresponding  gauge invariant
field strengths are thus
\begin{equation}
F=dA\ \Leftrightarrow\ F_{\mu_0\dots\mu_p}=(p+1)
\partial _{[\mu_0}A_{\mu_1\dots\mu_p]}\ ,
\end{equation}
and the standard
Lagrangian for a set of free $p$-form gauge fields
is a linear combination of Maxwell-type kinetic terms
$F_{\mu_0\dots\mu_p}F^{\mu_0\dots\mu_p}$.

A systematic investigation of
the possible interaction vertices
which can be added consistently to such a free Lagrangian
$L^{(0)}$ was carried out
by Henneaux and Knaepen \cite{bra-HK1,bra-HK2,bra-HK3}. 
They studied consistent deformations of the free
Lagrangian $L^{(0)}$ and of the gauge transformations
$\delta_\mathrm{gauge}^{(0)}$,
\begin{eqnarray}
L&=& L^{(0)}+g^\alpha V^{(1)}_\alpha+g^\alpha g^\beta
V^{(2)}_{\alpha\beta}+\dots
\\
\delta_\mathrm{gauge}&=&\delta_\mathrm{gauge}^{(0)}
+g^\alpha\delta_\mathrm{gauge}^{(1)}{}_\alpha
+g^\alpha g^\beta\delta_\mathrm{gauge}^{(2)}{}_{\alpha\beta}+\dots\ ,
\end{eqnarray}
where $g^\alpha$ are continuous coupling constants (deformation parameters),
such that the deformed Lagrangian $L$ is invariant under
the deformed gauge transformations $\delta_\mathrm{gauge}$ 
modulo a total derivative,
\begin{equation}
\delta_\mathrm{gauge}\, L=\partial _\mu K^\mu.
\label{bra-inv}
\end{equation}
To first order in the coupling constants, (\ref{bra-inv}) requires
that the
$V^{(1)}_\alpha$ be $\delta_\mathrm{gauge}^{(0)}$-invariant 
on-shell in the free theory 
modulo a total derivative. Furthermore, without loss of generality,
one may neglect all $V^{(1)}_\alpha$ which
vanish on-shell in the free theory 
modulo a total derivative because they can be removed by field 
redefinitions (such vertices are therefore called
trivial ones). Henneaux and Knaepen found the following result for 
the remaining first-order vertices:

{\em Category 1:} Vertices that are 
$\delta_\mathrm{gauge}^{(0)}$-invariant off-shell
modulo a total derivative and therefore
do not modify the gauge transformations to first order. 
There are
two types of such vertices (modulo total derivatives). 
Those of the first type 
depend on $p$-form gauge fields only via
the field strengths $F_{\mu_0\dots\mu_p}$ and their derivatives.
Of course, there are infinitely many vertices of this type.
Those of the second type are vertices of the
Chern-Simons type
\begin{equation}
A\wedge F\wedge\dots\wedge F
\label{bra-CS}
\end{equation}
where the $F$'s may have different form-degrees
and all form-degrees must sum up to the spacetime dimension.
These vertices are $\delta_\mathrm{gauge}^{(0)}$-invariant only modulo a
total derivative.

{\em Category 2:}
Vertices that are $\delta_\mathrm{gauge}^{(0)}$-invariant
only on-shell in the free theory modulo a total derivative. 
These vertices are of particular
interest because they are accompanied by deformations
of the gauge transformations.
A remarkable result is that, when ordinary
gauge fields (1-form gauge potentials) are absent, all these
vertices can be brought to the following form (modulo
trivial vertices and vertices of category 1):
\begin{equation}
A\wedge F\wedge\dots\wedge F\wedge 
\underbrace{{^*F}\wedge\dots\wedge{^*F}}_{\mbox{at\,least\,one\,${^*F}$}}
\label{bra-LHK}
\end{equation}
where ${^*F}$ denotes the Hodge dual of $F$ and there must be
at least one ${^*F}$ because otherwise the vertex would be
of the Chern-Simons type (\ref{bra-CS}).
Again, the $F$'s may have different form-degrees
and all form-degrees must sum up to the spacetime dimension.
Therefore there are only finitely many vertices (\ref{bra-LHK})
for a finite number of $p$-form gauge fields. The first order
deformations of the gauge transformations
which correspond to a vertex (\ref{bra-LHK}) take the
form
\begin{equation}
\delta_\mathrm{gauge}^{(1)} A
=\omega\wedge F\wedge\dots\wedge F\wedge 
{^*F}\wedge\dots\wedge{^*F}
\label{bra-deltaHK}
\end{equation} 
where one of the $^*F$'s that occurs in
(\ref{bra-LHK}) is omitted (for instance, when (\ref{bra-LHK}) contains
only one $^*F$, then (\ref{bra-deltaHK}) contains no $^*F$).
When 1-form gauge potentials are present, (\ref{bra-LHK})
still gives nontrivial first-order vertices of category 2, but
then there may be additional
vertices of category 2 which cannot be brought to the form
(\ref{bra-LHK}). In particular,
when at least three 1-form gauge potentials are present, there are
Yang-Mills cubic vertices which differ
from (\ref{bra-LHK}) because they contain two `naked' gauge potentials
instead of only one (the structure of
Yang-Mills cubic vertices
is $A\wedge A\wedge {^*F}$ where the $A$'s are
1-form gauge potentials and $F$ is a 2-form field strength).

In four-dimensional spacetime there are three different
types of cubic vertices (\ref{bra-LHK}) involving 1-form gauge
potentials $A_1$, 2-form potentials $A_2$ and corresponding
field strengths $F_2=dA_1$ and
$F_3=dA_2$:
\begin{eqnarray}
& A_2\wedge {{^*F}}_3\wedge {^*F}_3 &
\label{bra-FT1}\\
& A_1\wedge {^*F}_2\wedge {^*F}_3 &
\label{bra-HK1}\\
& A_1\wedge F_2\wedge {^*F}_3\ .&
\label{bra-CM1}\end{eqnarray}
These are the vertices mentioned in the introduction.

\section{Overview of the new models}\label{bra-overview}

In accordance with commonly used nomenclature (which is actually
somewhat unfair, see remarks at the end of this section),
the vertices (\ref{bra-FT1}), (\ref{bra-HK1}) and (\ref{bra-CM1})
will be referred to as
``Freedman-Townsend'' (FT), ``Henneaux-Knaepen'' (HK)
and ``Chapline-Manton'' (CM) vertices, respectively.
Each of the new supersymmetric models reviewed here contains
at least one of these vertices. We label 
1-form potentials and 2-form potentials by indices
$a=1,2,\dots$ and $i=1,2,\dots$ respectively,
and denote their component fields by 
$A_\mu^a$ and $B_{\mu\nu}^i=-B_{\nu\mu}^i$.
The field strengths of $A_\mu^a$
are denoted by $F_{\mu\nu}^a=\partial _\mu A_\nu^a-\partial _\nu A_\mu^a$,
the Hodge-dualized field strengths of $B_{\mu\nu}^i$
by $H^{i\mu}=\frac 12 \varepsilon^{\mu\nu\rho\sigma}
\partial _\nu B^i_{\rho\sigma}$. The vertices
(\ref{bra-FT1}), (\ref{bra-HK1}) and (\ref{bra-CM1}) read
explicitly, using a suitable normalization,
\begin{eqnarray}
\mbox{FT vertices:}&&\quad
\frac 14\,f_{ijk}\, H_\mu^i H_\nu^j 
B_{\rho\sigma}^k\,\varepsilon^{\mu\nu\rho\sigma}
\label{bra-FT}\\
\mbox{HK vertices:}&&\quad
T_{iab}\, H^i_\mu F^{a\mu\nu}A^b_\nu
\label{bra-HK}\\
\mbox{CM vertices:}&&\quad
\frac 12\,S_{iab}\, H^i_\mu F^a_{\nu\rho}A^b_\sigma\,
\varepsilon^{\mu\nu\rho\sigma}
\label{bra-CM}
\end{eqnarray}
where the $f_{ijk}$, $T_{iab}$ and $S_{iab}$ are
constant coefficients, with
\[
f_{ijk}=-f_{jik}\quad ,\quad S_{iab}=S_{iba}\ .
\]
[$S_{iab}=S_{iba}$ can be imposed without loss of generality because
$S_{i[ab]}$ can be removed from the vertices (\ref{bra-CM}) 
by subtracting trivial vertices.] 
These coefficients are subject to conditions
imposed by (\ref{bra-inv}) at second order in the coupling
constants (deformation parameters). 
Viewing $T_{iab}$ and $S_{iab}$ as the entries
of matrices $T_i$ and $S_i$, these conditions read
\begin{eqnarray}
& f_{ijl}f_{klm}+f_{jkl}f_{ilm}+f_{kil}f_{jlm}=0 &
\label{bra-jac}\\[4pt]
& [T_i,T_j]=f_{ijk}\,T_k &
\label{bra-rep}\\[4pt]
& (S_i T_j-S_j T_i)+(S_i T_j-S_j T_i)^\top=f_{ijk}\,S_k\ . &
\label{bra-new}
\end{eqnarray} 
To derive these conditions, it was assumed 
that the zeroth order Lagrangian is $L^{(0)}=-(1/2) H^{\mu i}H_\mu^i
-(1/4)F_{\mu\nu}^a F^{\mu\nu a}$, and that 
(\ref{bra-FT1}), (\ref{bra-HK1}) and (\ref{bra-CM1})
are the only vertices of category 2 with non-vanishing
coefficients (vertices of category 1
do not modify these conditions, but switching
on other vertices of category 2 might cause modifications
or lead to additional conditions).

(\ref{bra-jac}) and (\ref{bra-rep}) were already found
in \cite{bra-HK1} and require that
the $f_{ijk}$ be structure
constants of a Lie algebra and that the $T_i$ be representation
matrices of that Lie algebra, respectively. 
(\ref{bra-new}) was not derived in a previous work, to my knowledge.
It requires that the symmetric parts of the matrices
$2(S_i T_j-S_j T_i)$ be equal to $f_{ijk}\,S_k$. This is fulfilled,
for instance, if $S_i=N T_i+T_i^\top N$ where $N$ is an arbitrary
symmetric matrix (i.e.,
$S_{iab}=N_{ac}T_{icb}+N_{bc}T_{ica}$ with
$N_{ab}=N_{ba}$), but there are other solutions as well.

The corresponding first order deformations of the gauge transformations
are
\begin{eqnarray}
\delta^{(1)}_\mathrm{gauge}\, B_{\mu\nu}^i &=&
-f_{ijk} (H_\mu^j \omega^k_\nu-H_\nu^j \omega^k_\mu)
-\frac 12\, \varepsilon_{\mu\nu\rho\sigma} T_{iab} F^{\rho\sigma a}\omega^b
+S_{iab} F^a_{\mu\nu} \omega^b
\nonumber\\
\delta^{(1)}_\mathrm{gauge}\, A_\mu^a &=& - T_{iab} H_\mu^i \omega^b.
\end{eqnarray} 

The following table gives an overview of the new supersymmetric
models. The vertices discussed above are used to distinguish
between the various models. In addition the number of supersymmetries
(N=1 or N=2 supersymmetry) and the supersymmetry multiplets
are given. In the case of N=1 supersymmetry, T and V stand for 
tensor multiplets (also called linear multiplets) 
and vector multiplets
respectively. 
In the case of N=2 supersymmetry, VT, TT and V stand for
vector-tensor multiplets, double-tensor multiplets
and vector multiplets respectively.
\medskip

\begin{tabular}{l|l|l|l}
susy & multiplets & interactions & papers\\ 
\hline\rule{0em}{2.5ex}
N=2 & {VT}\,,\,{V} & {HK}\,,\,{CM} & 
\cite{bra-1,bra-2,bra-7,bra-8,bra-10}\\
\hline\rule{0em}{2.5ex}
N=2 & {VT}\,,\,{V} & {CM} & 
\cite{bra-3,bra-4,bra-5,bra-9}\\
\hline\rule{0em}{2.5ex}
N=2 & {VT} & {CM} & \cite{bra-6}\\
\hline\rule{0em}{2.5ex}
N=1 & {T}\,,\,{V} & 
{FT}\,,\,{HK}\,,\,{CM} & \cite{bra-11}\\
\hline\rule{0em}{2.5ex}
N=2 & {VT} & {HK} & \cite{bra-12}\\
\hline\rule{0em}{2.5ex}
N=2 & {TT}\,,\,{V} & {FT}\,,\,{HK} & \cite{bra-13}
\end{tabular}
\medskip

Of course, this table characterizes the various models
only very roughly. 
The example in the next section is to illustrate characteristic
features of these models. It is beyond the scope of this
paper to review the various models in 
greater detail but I would like to add
at least a few remarks: (a) Among all these models
only those in \cite{bra-7} are locally supersymmetric, the
other ones are globally supersymmetric.
(b) The works on the VT multiplet
overlap in part because some of these
works rederive models which had already been found
by means of other methods in previous works.
(c) Models in the same row of the table may of course still differ. 
For instance,
CM vertices in two models with the same multiplet content 
may contain different Chern-Simons forms
(in the literature,
this has led to a distinction between
``linear'' and ``nonlinear'' VT multiplets \cite{bra-2}).
Different CM couplings
correspond to different solutions
to Eq.\ (\ref{bra-new}). Of course, analogous statements apply to the
FT and HK vertices.
(d) Some of the models
in \cite{bra-11} possess extended ($N\geq 2$)
supersymmetry. For instance, 
it has been pointed out in \cite{bra-12} that the model constructed
there can be obtained from \cite{bra-11}. 
However, it is not clear how to sieve out systematically
those models in \cite{bra-11} which have extended supersymmetry.

Finally a few comments on the history may be in order.
Models with FT interactions were constructed already
by Ogievetsky and Polubarinov \cite{bra-OP} a long time
before the work by Freedman and Townsend \cite{bra-FT}.
CM interactions have a long history too. It seems that
they appeared first
in the early 80's \cite{bra-NT,bra-Bergshoeff,bra-CM} and, again,
the work by Chapline and Manton was not the first 
one with such interactions.
CM interactions attracted particular attention because of their
crucial r\^ole in the Green-Schwarz anomaly cancellation
mechanism \cite{bra-GS} (the anomaly cancellation
is made possible by the deformation
of the gauge transformations
associated with CM vertices, see section \ref{bra-sec2}).

HK interactions (in four-dimensional spacetime)
were discovered much later. However, the first models 
with such interactions were
not found by Henneaux and Knaepen. Rather,
it seems that HK interactions occurred for the first time in
\cite{bra-1} where the central charge of the VT multiplet
was gauged. The connection of that gauging to HK vertices 
is the following.
Gauging the central charge (e.g., via the Noether method) 
gives rise to
a vertex $V_\mu j^\mu$ where $V_\mu$ is
a 1-form gauge field and $j^\mu$ is the Noether current
corresponding to the central charge symmetry. 
That Noether current is $j^\mu=H_\nu F^{\nu\mu}$,
and thus the vertex $V_\mu j^\mu$ is a HK vertex.
Combined FT and HK interactions, and
the relation to Lie algebras, were found afterwards
by Henneaux and Knaepen \cite{bra-HK1}.
It seems that the first and so far only work with models
containing simultaneously FT, HK and CM vertices
is \cite{bra-11}.

\section{Example}\label{bra-example}

The example is an N=2 supersymmetric
model coupling one TT multiplet to two V multiplets
and involves HK vertices but no FT or CM vertices.
A TT multiplet contains two 2-form gauge potentials
$B_{\mu\nu}^i$ ($i=1,2)$, two real scalar fields $a^i$
and two Weyl fermions $\chi$ and $\psi$.
Each V multiplet contains a 1-form gauge potential
$A_\mu$, a complex scalar field $\phi$ and two
Weyl fermions $\lambda^i$. The V multiplets are
labeled by the index $a=1,2$. This
field content is supplemented with auxiliary fields $h_\mu^i$
which are embedded in the TT multiplet.
These auxiliary fields allow one to construct the model
in a compact polynomial form.
In fact, it would be very cumbersome to construct
the model without these auxiliary fields because
of the complicated nonpolynomial structure which arises then,
see below. 
Note that, in contrast to other supersymmetric
models, the auxiliary fields do not lead to an
off-shell closed supersymmetry algebra. On the contrary,
the auxiliary fields make the supersymmetry algebra
even ``more open'' (a formulation of the TT multiplet
with an off-shell closed supersymmetry algebra is not
known).

\[
\begin{array}{c|c|c}
& \mbox{bosons} & \mbox{Weyl-fermions} 
\\
\hline\rule{0em}{2.5ex}
\mbox{{TT}}& {B_{\mu\nu}^i}\quad a^i\quad 
({h_\mu^i}) &  \chi\quad \psi
\\
\hline\rule{0em}{2.5ex}
\mbox{{V$^a$}}& {A_\mu^a}\quad \phi^a &
\lambda^{ai}
\end{array}
\]

Thanks to the inclusion of the auxiliary fields,
the Lagrangian takes the following simple form
(using conventions as \cite{bra-WB} adapted to
the Minkowski metric $\mathrm{diag}(1,-1,-1,-1)$),
\begin{eqnarray}
L=
\partial _\mu a^i \partial ^\mu a^i+{h_\mu^i h^{\mu i}}
+2{h_\mu^i} H^{\mu i}-{\mathrm{i}} \chi\partial \bar \chi
-{\mathrm{i}} \psi\partial \bar \psi
\nonumber\\
-\frac 14 {\hat F^a_{\mu\nu}\hat F^{a\mu\nu}}
+\frac 12 {\hat D_\mu}\phi^a{\hat D^\mu} \bar \phi^a
-2{\mathrm{i}} \lambda^{ia}{\hat D} \bar \lambda^{ia}
\label{bra-L}
\end{eqnarray}
where
\begin{eqnarray*}
{\hat F^a_{\mu\nu}}&=&{\hat D_\mu} A_\nu^a-{\hat D_\nu} A_\mu^a
=\partial _\mu A_\nu^a+g^i {h_\mu^i}\varepsilon^{ab}A_\nu^b
-(\mu\!\leftrightarrow\!\nu)
\\
{\hat D_\mu}\phi^a&=&\partial _\mu\phi^a+g^i {h_\mu^i}\varepsilon^{ab}\phi^b
\\
{\hat D}\bar \lambda^{ia}&=&\sigma^\mu (\partial _\mu\bar \lambda^{ia}
+g^i {h_\mu^i}\varepsilon^{ab}\bar \lambda^{ib}).
\end{eqnarray*}
The $g^i$ are real coupling constants (deformation parameters). 
Note that ${\hat D_\mu}$ has the form of a covariant derivative
even though the auxiliary fields cannot be viewed
as gauge fields (in fact, they substitute for field strengths,
as the equations of motion give $h_\mu^i=-H_\mu^i+\dots$).
The auxiliary fields also simplify the structure of the gauge
and supersymmetry transformations considerably.
The gauge transformations read
\begin{eqnarray*}
\delta_\mathrm{gauge}\, A_\mu^a&=&
{\hat D_\mu}{\omega^a}=\partial _\mu {\omega^a}
+g^i {h_\mu^i}\varepsilon^{ab}{\omega^b}
\\
\delta_\mathrm{gauge}\, B_{\mu\nu}^i &=& 
\frac 14\, g^i {\omega^a} \varepsilon^{ab}\varepsilon_{\mu\nu\rho\sigma}
{\hat F^{b\rho\sigma}}
+\partial _\mu {\omega^i_\nu}-\partial _\nu {\omega^i_\mu}
\\
\delta_\mathrm{gauge} &=&0\quad\mbox{on other fields}
\end{eqnarray*}
where $\omega^a$ and $\omega_\mu^i$ are the gauge parameter
fields associated with $A^a_\mu$ and $B_{\mu\nu}^i$ respectively.
The supersymmetry transformations read, 
with constant anticommuting Weyl-spinors $\xi^i$ as transformation
parameters,
\begin{eqnarray*}
\delta_\mathrm{susy}\, A^a_\mu &=& 
\varepsilon^{ij}\xi^i\sigma_\mu \bar \lambda^{ja}
- \xi^i {\Gamma^{i}} \varepsilon^{ab} A^b_\mu+\mbox{c.c.}
\\
\delta_\mathrm{susy}\, \phi^a &=& 2\, \xi^i\lambda^{ia}
- (\xi^i {\Gamma^{i}}+\bar \xi^i {\bar \Gamma^{i}}) \varepsilon^{ab}\phi^b
\\
\delta_\mathrm{susy}\, \lambda^{ia} &=& \frac{{\mathrm{i}}}2\, 
(\varepsilon^{ij}
                   \xi^j \sigma^{\mu\nu}{\hat F^a_{\mu\nu}}
-\bar \xi^i\bar \sigma^\mu {\hat D_\mu}\phi^a)
-(\xi^j {\Gamma^j}+\bar \xi^j \bar \Gamma^j) \varepsilon^{ab} \lambda^{ib}
\\
\delta_\mathrm{susy}\, B_{\mu\nu}^i&=&
-\varepsilon^{ij}\xi^j\sigma_{\mu\nu}\chi+\xi^i\sigma_{\mu\nu}\psi
\\
&&+{\mathrm{i}} g^i\varepsilon^{ab}(\bar \phi^a\xi^j\sigma_{\mu\nu}\lambda^{jb}
+\varepsilon^{jk}A_{[\mu}^a\xi^j\sigma^{}_{\nu]}\bar \lambda^{kb})
+\mbox{c.c.}
\\
\delta_\mathrm{susy}\, a^i&=&
\frac 12\, (\xi^i\chi-\varepsilon^{ij}\xi^j\psi)+\mbox{c.c.}
\\
\delta_\mathrm{susy}\,\chi&=&-\bar \xi^i\bar \sigma^\mu(
\varepsilon^{ij}{h_\mu^j}+{\mathrm{i}}\, \partial _\mu a^i)
\\
\delta_\mathrm{susy}\,\psi&=&-\bar \xi^i\bar \sigma^\mu(
{h_\mu^i}+{\mathrm{i}}\, \varepsilon^{ij}\partial _\mu a^j)
\\
\delta_\mathrm{susy}\, {h_\mu^i}&=&
\frac{{\mathrm{i}}}2\, \partial _\mu (\xi^i\psi-\varepsilon^{ij}\xi^j\chi)
+\mbox{c.c.}
\end{eqnarray*}
where
\[
{\Gamma^i}=\frac {{\mathrm{i}}}2\,  
{g^j}(\varepsilon^{ij}\chi+\delta^{ij}\psi).
\]
The commutator
algebra of the supersymmetry and gauge transformations is
rather complicated off-shell but on-shell it is quite simple,
\begin{eqnarray}
[\delta^{}_\mathrm{susy},\delta'_\mathrm{susy}]
&\approx&\delta_\mathrm{translation}+\delta_\mathrm{gauge}
\label{bra-alg1}\\
{}[\delta_\mathrm{susy},\delta_\mathrm{gauge}]&\approx& 
\delta'_\mathrm{gauge}
\label{bra-alg2}\\
{}[\delta^{}_\mathrm{gauge},\delta'_\mathrm{gauge}]&\approx& 0,
\label{bra-alg3}
\end{eqnarray}
where $\approx$ is equality on-shell. (\ref{bra-alg1}) is the
standard N=2 supersymmetry algebra on-shell
(modulo gauge transformations),
with vanishing central charge.
I remark that the gauge transformations which appear on the
right hand side of (\ref{bra-alg1}) involve explicitly the
spacetime coordinates, see \cite{bra-13} and \cite{bra-hidden}
for details and comments on this point. (\ref{bra-alg2}) illustrates
a feature typical of many of the new models, namely that
gauge and supersymmetry transformations do not commute (not even
on-shell). Explicitly, the gauge parameter fields $\omega^{a\prime}$
and $\omega^{i\prime}_\mu$
of $\delta'_\mathrm{gauge}$
on the right hand side of (\ref{bra-alg2}) read
\begin{eqnarray*}
\omega^{a\prime}&=&(\xi^i {\Gamma^{i}}+\bar \xi^i {\bar \Gamma^{i}}) 
\varepsilon^{ab}\omega^b
\\
\omega^{i\prime}_\mu&=&-\frac{{\mathrm{i}}}2\,g^i\varepsilon^{ab}
\varepsilon^{jk}\omega^a
(\xi^j\sigma_\mu \bar \lambda^{kb}-\lambda^{kb}\sigma_\mu\bar \xi^j)
\end{eqnarray*}
where the $\xi$'s and $\omega$'s are supersymmetry parameters
and gauge parameter fields of $\delta_\mathrm{susy}$
and $\delta_\mathrm{gauge}$
on the left hand side of (\ref{bra-alg2}).
According to (\ref{bra-alg3}), the gauge transformations commute on-shell
which is also typical of the new models [note: the algebra of
the gauge transformations is not related to the Lie algebra
underlying Eqs.\ (\ref{bra-jac}) through (\ref{bra-new})!].

Let me finally discuss the nonpolynomial structure
which arises when one
eliminates the auxiliary fields.
The Lagrangian (\ref{bra-L}) contains the auxiliary fields at most
quadratically,
\begin{eqnarray*}
L&=&-\frac 14 F^a_{\mu\nu}F^{a\mu\nu}
+\partial _\mu a^i \partial ^\mu a^i
+\frac 12 \partial _\mu\phi^a\partial ^\mu \bar \phi^a
\\
&&-{\mathrm{i}} \chi\partial \bar \chi-{\mathrm{i}}\psi\partial \bar \psi
-2{\mathrm{i}}\lambda^{ia}\partial  \bar \lambda^{ia}
+2{h_\mu^i} {\cal H}^{\mu i}
+{h_\mu^i}K^{\mu i,\nu j} {h_\nu^j}
\end{eqnarray*}
where
\begin{eqnarray*}
&{\cal H}^{\mu i}=H^{\mu i}
-{g^i}\varepsilon^{ab}(\frac 12 F^{a\mu\nu}A_\nu^b 
+\frac 14\phi^a \stackrel{\!\leftrightarrow\!}{\partial ^\mu}\bar \phi^b
+{\mathrm{i}} \lambda^{ja}\sigma^\mu\bar \lambda^{jb})&
\\[6pt]
&K^{\mu i,\nu j}=
\eta^{\mu\nu}\delta^{ij}+\frac 12 {g^i g^j}
[\eta^{\mu\nu}(\phi^a\bar \phi^a-A_\rho^a A^{a\rho})+A^{a\mu}A^{a\nu}]&
\end{eqnarray*}
The auxilary fields can be eliminated by solving their
algebraic equations of motion. The solution is
\begin{equation}
{h_\mu^i}=-(K^{-1})_{\mu i,\nu j}{\cal H}^{\nu j},
\label{bra-aux}
\end{equation}
where $K^{-1}$ is the inverse of the field dependent matrix $K$,
$(K^{-1})_{\mu i,\rho k}K^{\rho k,\nu j}=\delta^\nu_\mu\delta^j_i$.
Note that $K$ does not involve derivatives of the fields and therefore
$K^{-1}$ is nonpolynomial in the fields but still local.
Hence, using (\ref{bra-aux}),
the Lagrangian, gauge and supersymmetry transformations become
nonpolynomial but remain strictly local. Expanding the
resulting Lagrangian in the coupling constants, one finds at
first order HK vertices as well as vertices of category 1
which complete the HK vertices such that the sum is supersymmetric
on-shell in the free theory modulo a total derivative,
\begin{eqnarray}
L&=&-\frac 14 F^a_{\mu\nu}F^{a\mu\nu}
+\partial _\mu a^i \partial ^\mu a^i
+\frac 12 \partial _\mu\phi^a\partial ^\mu \bar \phi^a
\nonumber\\
&&-{\mathrm{i}} \chi\partial \bar \chi-{\mathrm{i}}\psi\partial \bar \psi
-2{\mathrm{i}}\lambda^{ia}\partial  \bar \lambda^{ia}
-{\cal H}^{\mu i}\,(K^{-1})_{\mu i,\nu j}\, {\cal H}^{\nu j}
\nonumber\\
&=&
L^{(0)}+\underbrace{
{g^i}\varepsilon^{ab}H_{\mu}^iF^{a\mu\nu}A_\nu^b}_{\mbox{HK vertices}}
+\underbrace{{g^i}\varepsilon^{ab}H_{\mu}^i(\frac 12\phi^a 
\stackrel{\!\leftrightarrow\!}{\partial ^\mu}\bar \phi^b
+2{\mathrm{i}} \lambda^{ja}\sigma^\mu
\bar \lambda^{jb})}_{\mbox{category 1 vertices}
\atop
\mbox{(susy completion of HK vertices)}}+\dots\quad
\label{bra-exp}\end{eqnarray}

It was mentioned already that nonpolynomial structures as in this
example are typical of the new gauge theories. 
They cannot be avoided in models with
FT or HK vertices because they are
necessary consequences of these vertices, already in the
non-supersymmetric case. The use of appropriate auxiliary
fields that simplify the construction is an almost
indispensable tool for constructing complicated models of this type, 
especially supersymmetric ones. The finding of such
auxiliary fields and their embedding in supersymmetry multiplets
is in general a nontrivial and subtle
ingredient of the construction.
In contrast, models which contain
CM vertices but no FT or HK vertices are simpler
and the issue of auxiliary fields is less involved. In particular,
such models are not necessarily nonpolynomial although
supersymmetry often enforces a nonpolynomial dependence
on scalar fields even in such models.

\section{Comments}\label{bra-comments}

The following is a selection of open problems which may point
to possible further developments in the field:

(i) In my opinion,
the r\^ole of the matter fields (scalar fields, fermions) 
in the new supersymmetric models
has not been fully understood yet.
In particular, the relation of scalar fields to the
underlying geometry (Lie algebra)
is somewhat mysterious. A better understanding of this issue
might be a key to a deeper understanding of the supersymmetry
structure of the models and to a more systematic construction
of such models.

(ii) Systematic classifications of the possible
consistent and supersymmetric interactions involving
$p$-form gauge potentials, analogous to
the classification \cite{bra-HK1,bra-HK2,bra-HK3}
of non-supersymmetric interactions,
are largely missing.
An exception is the classification of the lowest dimensional
interaction vertices involving a TT multiplet in \cite{bra-13}.
Supersymmetry supplements (\ref{bra-inv}) with the 
additional requirement
$\delta_\mathrm{susy}L=\partial _\mu M^\mu$ where $\delta_\mathrm{susy}$
are the deformed supersymmetry transformations.
This restricts the possible
interactions as compared to the non-supersymmetric case, 
and relates coefficients of various interaction terms.
A typical example is (\ref{bra-exp}) where the coefficients of the
HK vertices are related to coefficients of interaction
vertices of category 1.
In fact, supersymmetry can even completely forbid interactions which
would be allowed if supersymmetry were not imposed.
An example is the absence of N=2 supersymmetric
CM couplings of the TT multiplet
\cite{bra-13}. Furthermore, it depends
on the supersymmetry multiplet structure which
interactions are possible. For instance, 
it was just mentioned that there are no N=2 supersymmetric
CM couplings involving the TT multiplet, whereas such couplings
do exist for the VT multiplet (cf.\ table in section \ref{bra-overview}).
Such results could be relevant in the context
of string theory when comparing
properties of different superstring vacua.

(iii) Locally supersymmetric models with FT or HK couplings
are almost completely missing so far. In fact, the only
exception is the work \cite{bra-7} where N=2 supergravity
models with VT multiplets were constructed.
The construction of locally supersymmetric extensions
of some of the other models could be of interest 
in the string theory context. In particular this
applies to supergravity models with the TT multiplet because of
the conjectured importance of this multiplet to
type IIB superstring vacua (cf.\ introduction).

(iv) Recall that FT, HK and CM vertices are special
cases of vertices (\ref{bra-LHK}). Non-supersymmetric
models in spacetime dimensions $>4$ with such vertices
have been constructed already \cite{bra-HK1,bra-BST}. Analogous
globally or locally supersymmetric models in
higher spacetime dimensions have not been constructed
so far. In fact it seems that the only vertices  (\ref{bra-LHK}) 
which have been used in supersymmetric models in
spacetime dimensions $>4$ so far are the familiar
CM vertices (\ref{bra-CM}). For instance, these vertices
occur in 10-dimensional supergravity in connection with
the Green-Schwarz anomaly cancellation mechanism
(cf.\ remarks at the end of section \ref{bra-overview}).

\end{document}